\documentstyle[sprocl,epsf]{article}

\begin{document}

\hfill TPI-MINN-99/30, HEP-TH-1804

\bigskip

\title{ HOW PENGUINS STARTED TO FLY\\~\\The
1999 Sakurai Prize Lecture\,\footnote{Talk at the 1999 Centennial Meeting of the
American Physical Society, March 20-26, on the occasion of receiving the 1999
Sakurai Prize for Theoretical Particle Physics.}}
\author{ARKADY  VAINSHTEIN}
\address{Theoretical Physics Institute, University of Minnesota,
 Minneapolis, MN 55455}

\maketitle

\abstracts{
A mechanism explaining a strong enhancement of nonleptonic weak decays was
suggested in 1975, later to be dubbed the penguin. This mechanism extends
Wilson's ideas about the operator product expansion at short distances  and reveals
an intricate interplay of subtle features of the theory such as heavy quark masses
in Glashow-Iliopoulos-Maini cancellation, light quarks shaping the chiral properties
of QCD, etc. The penguins have subsequently evolved to play a role in 
a variety of fields in present-day particle phenomenology. I will describe the
history of this idea and review its subsequent development. The recent
measurement of direct CP violation in $K$ decays gives a new confirmation of the
penguin mechanism.
 }

\section{History of the idea}
It was an exciting period, with Quantum Chromodynamics (QCD) emerging as {\em
the} theory of strong interactions,  when three of us -- Valya Zakharov, Misha
Shifman and I -- started in 1973 to work on QCD effects in weak processes. The
most dramatic signature of strong interactions in these processes is the so called 
$\Delta I=1/2$ rule in nonleptonic weak decays of strange particles.  Let me remind
you what this rule means by presenting the experimental value for the ratio of the
widths of $K_S\to\pi^+\pi^-$ and $K^+\to\pi^+\pi^0$ decays
\begin{equation}
\frac{\Gamma(K_S\to \pi^+\pi^-)}{\Gamma(K^+\to\pi^+\pi^0)}=450~.
\label{ratio}
\end{equation}
The isotopic spin $I$ of hadronic states is changed by 1/2 in the  $K_S\to\pi^+\pi^-$
weak transition and by 3/2 in  $K^+\to\pi^+\pi^0$, so  the $\Delta I=1/2$ 
dominance is evident.  

What does theory predict? The weak interaction has a
current$\times$current form. Based on this, Julian Schwinger suggested\,\cite{Sch} 
to estimate nonleptonic amplitudes as a product of matrix elements of currents,
i.e.\ as a product of  semileptonic amplitudes. This approximation, which implies
that the strong interaction does not affect the form of the weak nonleptonic
interaction,  gives 9/4 for the ratio (\ref{ratio}). Thus, the theory is off by a factor
of two hundred! We see that strong interactions crucially affect nonleptonic weak
transitions.

The conceptual explanation was suggested\,\cite{KW} by Kenneth
Wilson  in the context of the Operator Product Expansion (OPE) which he
introduced\,\footnote{In Russia we had something of a preview of the OPE ideas
worked out by Sasha Patashinsky and Valery Pokrovsky (see their book\,\cite{PP})
in applications to phase transitions and by Sasha Polyakov in field theories.}.
Assuming scaling for OPE coefficients at short distances, Wilson related the 
enhancement of the $\Delta I=1/2$ part of the interaction with its more singular
behavior at short distances as compared with the $\Delta I=3/2$ part.  In the
pre-QCD era it was difficult to test this idea having no real theory of the strong
interaction.  With  the advent of QCD all this changed. The phenomenon of
asymptotic freedom gives  full theoretical control of short distances. Note in
passing that the American Physical Society also followed this development: the
discoverers of asymptotic freedom, Gross, Politzer, and Wilczek, became
recipients of the 1986 Sakurai Prize.

In QCD  the notion of OPE in application to nonleptonic weak interactions can be
quantified in more technical terms as one can calculate the effective Hamiltonian for
weak transitions at short distances. The weak interactions are carried  by $W$
bosons, so the characteristic distances are $\sim 1/m_W$, with $m_W=80$~GeV.
The QCD analysis at these distances  in the effective Hamiltonian was done in 1974
by Mary~K.~Gaillard with Ben Lee\,\cite{GL}, and by Guido Altarelli with Luciano
Miani\,\cite{AM}. Asymptotic freedom at short distances means that the strong 
interaction effects have a logarithmic dependence on momentum rather
than power-like behavior, as was assumed in Wilson's original analysis. The
theoretical parameter determining the effect  is $\log(m_W/\Lambda_{\rm QCD})$
where
$\Lambda_{\rm QCD}$ is a hadronic scale. 
These pioneering works  brought both good and bad news. The good  news
was that, indeed, strong interactions at short distances logarithmically enhance the
$\Delta I=1/2$ transitions and suppress the $\Delta I=3/2$ ones. The bad news
was that quantitatively the effect fell short  of an explanation of the ratio (1).

Besides $1/m_W$ and $1/\Lambda_{\rm QCD}$ there are scales
provided by masses of heavy quarks $t$, $b$ and $c$. In 1975 the object
of our study was  distances of order $1/m_c$ -- the top and bottom quarks were
not yet discovered. Introduction of top and bottom quarks practically does not
affect the 
$\Delta S=1$ nonleptonic transitions. However, it is different with charm.  At first
sight, the
$c$ quark loops looked to be  unimportant for nonleptonic decays of strange
particles in view of the famous Glashow-Illiopoulos-Miani cancellation\,\cite{GIM}
(GIM) with corresponding  up quark loops. In 1975  the
belief that this  cancellation produced the suppression factor
$$
\frac{m_c^2-m_u^2}{m_W^2}
$$ 
was universal, which is the reason why the effect of heavy quarks was overlooked.
We found instead that:
\begin{itemize}
\item[\bf (i)] The cancellation is distance dependent. Denoting $r=1/\mu$, we have
\begin{eqnarray}
&& \frac{m_c^2-m_u^2}{\mu^2}\,,\qquad\qquad{\rm for}~m_c\ll \mu\le m_W\,;
\nonumber\\[0.1cm]
&& \log\frac{m_c^2}{\mu^2}\,,\qquad\qquad~~~~{\rm for}~\mu\ll m_c\,.
\end{eqnarray}
{\em No suppression} below $m_c$!
\item[\bf (ii)] 
Moreover, new operators
appearing in the effective Hamiltonian at distances larger than $1/m_c$ are
qualitatively different -- they contain right-handed light quark fields in contrast
to the purely left-handed structures at distances much smaller than $1/m_c$ (see
the next section for their explicit form).  It was surprising that right-handed
quarks become strongly involved in weak interactions in the Standard Model 
with its left-handed weak currents. The right-handed quarks are coupled
via gluons which carry no isospin; for this reason new operators contribute to 
$\Delta I=1/2$ transitions only. 

\item[\bf (iii)]
For the mechanism we suggested it was crucial 
that  the matrix elements of novel operators were much larger than those for
purely left-handed operators. The enhancement  appears via the ratio
$$
\frac{m_\pi^2}{m_u +m_d}\sim 2 \,{\rm GeV}\,,
$$
which is large due to the small light
quark masses. The small values of these masses was a new idea at the time,
advocated in 1974 by Heiri Leutwyler\,\cite{Leut} and Murray
Gell-Mann\,\cite{MGM}. 
The origin of this large scale is not clear to this day  but it shows that in the
world of light hadrons there is, besides the evident momentum scale
$\Lambda_{\rm QCD}$, some other scale, numerically much  larger. 
\end{itemize}

Thus,  the explanation of the $\Delta I=1/2$ enhancement comes as a
nontrivial interplay of OPE, GIM cancellation, the heavy quark scale, and
different intrinsic scales in light hadrons. I will discuss the construction in  more
detail below. However,  I will first digress  to explain what the mechanism we
suggested has in common with penguins.

We had a hard time communicating our idea to the world.
Our first publication was a short letter published on July 20, 1975 in the Letters to
the Journal of Theoretical and Experimental Physics. Although  an
English translation of JETP Letters was available in the West we sent a more
detailed version to  Nuclear Physics shortly after. What  happened then was a long
fight  for  publication; we answered numerous referee reports -- our paper
was considered by quite a number of experts. The main obstacle for referees was to
overcome their conviction about the GIM suppression  $(m_c^2 -m_u^2)/m_W^2$
and to realize that there is no such suppression  at distances
larger than $1/m_c$. Probably our presentation was too concise for them to follow. 
 
Eventually the paper was published
in the March 1977 issue of Nuclear Physics without any revision, but only  after we
appealed to David Gross who was then on the editorial board.   The process
took more than a year and a half! We were so  exhausted  by this fight that we
decided to send our next publication containing a detailed theory of nonleptonic
decays to Soviet Physics JETP instead of an international journal. Lev Okun
negotiated  a special deal for us with the editor Evgenii Lifshitz  to submit the paper
of a size almost twice the existing limit  in the journal -- paper was in  short
supply as was almost everything in the USSR. We paid our price: the paper was
published in 1977, but even years later many theorists referred to our preprints of
the paper without mentioning the journal publication.

On a personal note, let me mention that a significant part of the JETP paper was
done over the phone line -- Valya and Misha worked in ITEP, Moscow, while I was
in the Budker Institute of Nuclear Physics, Novosibirsk. The phone connection was
not very good and we paid terrific phone bills out of our own pockets. 

I think, that in recognition of our work in the world at large it was Mary K. Gaillard
who first broke the ice  -- she mentioned the idea in one
of her review talks. Moreover, she collaborated with John Ellis, Dimitri Nanopoulos,
and Serge Rudaz in the work\,\cite{EGNR} in which they applied a similar
mechanism to B physics.  It is in this work that the mechanism was christened  the
penguin.  

How come?
Figure 1 shows the key Feynman diagram for the new operators in the
form  we drew it in our original publications~\cite{VZS}. 
\begin{figure}[h]
\epsfxsize=8cm
\centerline{\epsfbox{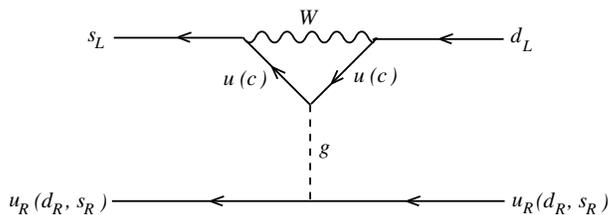}}\vspace*{-0.2cm}
\caption{Appearance of new operators due to quark loops}
\label{osix}
\end{figure}
It does not look at all penguin-like, right? Now look how a similar diagram is 
drawn in the paper of the four authors mentioned above.


\begin{figure}[h]
\epsfxsize=4cm
\centerline{\epsfbox{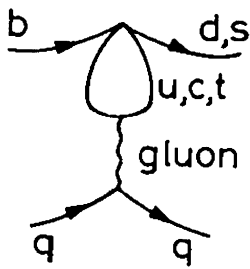}}
\vspace*{-0.4cm}
\caption{Quark loops in B decays}
\label{bpeng}
\end{figure}

\noindent
You see that that some measures were taken to make the diagram reminiscent of a
penguin. Let me refer here to John Ellis' recollections\,\footnote{John sent his 
recollections to Misha Shifman in 1995,  who published them in the 
preface\,\cite{MS} to his book.} on how it happened:
\begin{quotation}
``   Mary K, Dimitri and I first got interested in what are now called
penguin diagrams while we were studying CP violation in the Standard
Model in 1976 ...  The penguin name came in 1977, as follows.
    
In the spring of 1977, Mike Chanowitz, Mary K and I wrote a paper on
GUTs predicting the $b$ quark mass before it was found. When it was found
a few weeks later, Mary K, Dimitri, Serge Rudaz and I immediately started
working on its phenomenology. That summer, there was a student at CERN,
Melissa Franklin who is now an experimentalist at Harvard. One evening,
she, I and Serge went to a pub, and she and I started a game of darts. We
made a bet that if I lost I had to put the word penguin into my next paper.
She actually left the darts game before the end, and was replaced by Serge,
who beat me. Nevertheless, I felt obligated to carry out the conditions of the
bet.

    For some time, it was not clear to me how to get the word into this $b$
quark paper that we were writing at the time. Then, one evening, after
working at CERN, I stopped on my way back to my apartment to visit some
friends living in Meyrin, where I smoked some illegal substance. Later,
when I got back to my apartment and continued working on our paper, I had
a sudden flash that the famous diagrams look like penguins. So we put the
name into our paper, and the rest, as they say, is history.''
\end{quotation}

I learned some extra details of the story from Serge Rudaz  who is my Minnesota
colleague now. He recollects that for him to beat John in darts was a miraculous
event. John was a very strong player and had his own set of darts which he brought
to the pub.

\section{Effective Hamiltonian}
Application of Wilson's OPE to nonleptonic $\Delta S=1$ decays means the
construction of the effective Hamiltonian $H^{\rm eff}$ as a sum over local
operators
${\cal O}_i$,
\begin{equation}
H^{\rm eff}(\mu)=\sqrt{2} \, G_F V_{us}^* V_{ud} \sum_i c_i (\mu)\,{\cal
O}_i(\mu)\,. 
\end{equation}
Here $V_{us}$,  $V_{ud}$ are elements of the Cabibbo-Kobayashi-Maskawa
mixing matrix and 
 $\mu$ denotes the so called normalization point, which is the inverse of the
shortest distance for which the effective Hamiltonian is to be applied, ${\cal O}_i$
are gauge invariant local operators made out of quark and gluon fields, and $c_i$
are  OPE coefficients ($c$ numbers).  

At distances larger than $1/m_c$, i.e.\ at
$\mu<m_c$, only light
$u,d,s$ quarks and gluons remain as  building material for the ${\cal O}_i$.
Operators can be ordered according to their canonical dimension $d$,
and the corresponding  OPE coefficients are proportional to $(1/m_W)^{d-4}$.
As a selection criterion for operators we used their transformation features in the
limit of chiral SU(2)$_L$$\times$SU(2)$_R$ symmetry, picking up operators which
are SU(2)$_R$ singlets. Under this criterion, the operator of lowest
dimension ($d=5$) is of gluomagnetic type (magnetic penguins),
\begin{equation}
T=i\bar s_R \sigma_{\mu\nu} t^a  d_L\,G_{\mu\nu}^a\,,
\label{opT}
\end{equation}
where $G_{\mu\nu}^a$ is the gluon field strength tensor, and $t^a~(a=1,\dots 8)$ are
3$\times$3 generators of the color SU(3). The corresponding OPE coefficient is
proportional to the strange quark mass $m_s$. For this reason the magnetic
penguins turn out to be not important in $\Delta S=1$ transitions. They are
important, however,  for the
$b$ quark whose mass is large. We will return to this point later.

The operator basis of
SU(3)$_R$ invariant operators of
$d=6$ consists of
 six four-fermion operators.  The first four operators are constructed
from left-handed quarks (and their antiparticles, which are right-handed),
\begin{eqnarray}
{\cal O}_1&=&\bar s_L \gamma_\mu d_L \,\bar u_L \gamma^\mu u_L-
\bar s_L \gamma_\mu u_L\, \bar u_L \gamma^\mu d_L,
\qquad~~~~\,({\bf
8_f},~\Delta I=1/2), \nonumber\\[0.15cm]
{\cal O}_2&=&\bar s_L \gamma_\mu d_L \,\bar u_L \gamma^\mu u_L+
\bar s_L \gamma_\mu u_L \,\bar u_L \gamma^\mu d_L +2\bar s_L
\gamma_\mu d_L \,\bar d_L \gamma^\mu d_L\nonumber\\[0.15cm]
&&+2\bar s_L \gamma_\mu d_L \,\bar s_L \gamma^\mu s_L,
\qquad\qquad\qquad\qquad\qquad~~({\bf 8_d},~\Delta I=1/2),
\nonumber\\[0.15cm] 
{\cal O}_3&=&\bar s_L \gamma_\mu d_L \,\bar u_L
\gamma^\mu u_L+
\bar s_L \gamma_\mu u_L \,\bar u_L \gamma^\mu d_L +2\bar s_L \gamma_\mu
d_L \,\bar d_L \gamma^\mu d_L\nonumber\\[0.15cm]
&&-3\bar s_L \gamma_\mu d_L \,\bar s_L\gamma^\mu s_L,
\qquad\qquad\qquad\qquad\qquad~~({\bf 27},~\Delta I=1/2),
\nonumber\\[0.15cm]
{\cal O}_4&=&\bar s_L \gamma_\mu d_L \,\bar u_L \gamma^\mu
u_L+
\bar s_L \gamma_\mu u_L \,\bar u_L \gamma^\mu d_L \nonumber\\[0.15cm]
&&-\bar s_L \gamma_\mu d_L \,\bar d_L \gamma^\mu d_L,
\qquad\qquad\qquad\qquad\qquad~~~\,({\bf 27},~\Delta I=3/2)\,.
\end{eqnarray}
Every quark field is a color triplet $q^i$ and summation over color
indices is implied, $\bar q_2\gamma^\mu q_1= (\bar q_2)_i\,\gamma^\mu (q_1)^i$.
What is marked in the brackets are the SU(3) and isospin features of the operators.

Two more four-fermion operators entering the set contain also right-handed quarks
(in SU(3)$_R$ singlet form),
\begin{eqnarray}
{\cal O}_5\!\!\!\!&=&\!\!\!\!\bar s_L \gamma_\mu t^a d_L \left(\bar u_R
\gamma^\mu t^a u_R+
\bar d_R \gamma^\mu t^a d_R+\bar s_R \gamma^\mu t^a s_R\right),
\quad\!\!\!({\bf 8},~\Delta I=1/2), \nonumber\\[0.15cm]
{\cal O}_6\!\!\!\!&=&\!\!\!\!\bar s_L \gamma_\mu  d_L \left(\bar u_R
\gamma^\mu  u_R+
\bar d_R \gamma^\mu  d_R+\bar s_R \gamma^\mu  s_R\right),
\qquad\qquad({\bf 8},~\Delta I=1/2). 
\end{eqnarray}
Operators ${\cal O}_5$ and ${\cal O}_6$ are different by color flow only.

The operator basis we have introduced is recognized now (some doubts were
expressed in the literature at the beginning). The standard set\,\cite{GW}
used presents some linear combinations of ${\cal O}_{1-6}$. Actually, the set is
five instead of six combinations, the completness of the basis was lost on the
way, although it is not important within the Standard Model.

\subsection{Evolution}

The effective Hamiltonian (2) may remind the reader of the Fermi theory 
of beta decay with its numerous  variants for four-fermion operators. While in
many respects the analogy makes sense, the difference is that the standard model
together with QCD allows us to fix  all coefficients $c_i$. In the leading logarithmic
approximation the evolution of the effective Hamiltonian at
$\mu>m_c$ was found in Refs.\,\cite{GL,AM}. Penguins do not appear in this 
range
and the result for $H^{\rm eff}(m_c)$ has a simple form:
\begin{equation}
\left(\begin{array}{c}
c_1(m_c)\\[0.2cm] c_2(m_c)\\[0.2cm] c_3(m_c)\\[0.2cm] c_4(m_c)\\[0.2cm]
c_5(m_c)\\[0.2cm] c_6(m_c)
\end{array}\right) 
\ =\
\left(\begin{array}{c}
-\left[\frac{\alpha_S(m_b)}{\alpha_S(m_W)}\right]^{4/b_5}
\left[\frac{\alpha_S(m_c)}{\alpha_S(m_b)}\right]^{4/b_4}\\[0.2cm]
\frac{1}{5}\left[\frac{\alpha_S(m_b)}{\alpha_S(m_W)}\right]^{-2/b_5}
\left[\frac{\alpha_S(m_c)}{\alpha_S(m_b)}\right]^{-2/b_4}\\[0.2cm]
\frac{2}{15}\left[\frac{\alpha_S(m_b)}{\alpha_S(m_W)}\right]^{-2/b_5}
\left[\frac{\alpha_S(m_c)}{\alpha_S(m_b)}\right]^{-2/b_4}\\[0.2cm]
\frac{2}{3}\left[\frac{\alpha_S(m_b)}{\alpha_S(m_W)}\right]^{-2/b_5}
\left[\frac{\alpha_S(m_c)}{\alpha_S(m_b)}\right]^{-2/b_4}\\[0.2cm]
0\\[0.2cm] 0
\end{array}\right) \,,
\end{equation}
where $\alpha_S(\mu)$ is the running coupling 
\begin{equation}
\alpha_S(\mu)=\frac{\alpha_S(\mu_0)}{1+b_N \frac{\alpha_S(\mu_0)}{2
\pi}\ln\frac{\mu}{\mu_0}}\,, \qquad b_N=11-\frac 23\, N
\end{equation}
in the range with $N$ ``active'' flavors.  The  modification  due to the $b$ quark is
not significant, a few percent numerically, and the $t$ quark effects do not appear
in the leading logarithmic approximation.
 
Penguin operators ${\cal O}_{5,6}$ show up due to evolution at 
$\mu$ below $m_c$,
\begin{equation}
c_i (\mu)=\left[\exp\left\{
\frac{\rho}{b_3}\ln\frac{\alpha_S(\mu)}{\alpha_S(m_c)}\right\}\right]_{ij}
c_j(m_c)\,,
\end{equation}
where the anomalous dimension matrix $\rho$ is
\begin{equation}
\rho=\left( \begin{array}{cccccc}
34/9&10/9&0&0&4/3&0\\
1/9&-23/9~&0&0&-2/3~&0\\
0&0&-2~&0&0&0\\
0&0&0&-2~&0&0\\
1/6&-5/6~&0&0&6& 3/2\\
0&0&0&0&16/3&0
\end{array}\right)\,.
\end{equation}

Numerical results for the OPE coefficients depend on the normalization point $\mu$,
 pushing $\mu$ as low as possible maximizes the effect of the evolution. We chose
the lowest $\mu$ as the point where $\alpha_S(\mu)=1$.  The values of
$c_{1,2,3,4}$ are relatively stable, say,  under variation of the $c$ quark mass but
the penguin coefficients $c_{5,6}$ depend on it rather strongly.  This is not
surprising, of course, since penguins are generated in the interval of virtual
momenta  between
$\mu$ and $m_c$. Numerically, even for $\mu$ as low as 200 MeV, the  coefficients
$c_{5,6}$ are rather small. Our 1975 estimates for $c_5$ were in the interval 
\begin{equation}
c_5=0.06\div0.14\,.
\end{equation}
With the present day coupling $\alpha_S(m_Z)=0.115$, it would be about
twice smaller. The values of the coefficients $c_{5,6}$ are small and unstable. We
will return to the problem of OPE coefficients at the low normalization point in
connection with the procedure of calculating matrix elements, essentially it is about
matching for perturbative and nonperturbative effects.

We also found the coefficient for the magnetic penguin operator~(\ref{opT}).
Two-loop calculations were necessary for the purpose. Nowadays due to the efforts
of few groups the next to leading  approximation has been found for all types of
weak processes, see, e.g.,  the review\,\cite{BBL}. Although it is nice to have an
accurate values of the OPE coefficients the main phenomenological effects come
from matrix elements as we will see in the next Section. The theoretical uncertainty
is much bigger there.

\section{Matrix elements and phenomenology of nonleptonic decays}

We used the naive quark model  to find matrix elements
of four-fermion operators ${\cal O}_{1-6}$~.  The model implies a factorization
for amplitudes  of $K$ mesons decays, in the hyperon decays it is also the case for
all operators but ${\cal O}_1$.  It is clear that the factorization does not work at the
range of $\mu$ where the strong coupling $\alpha_S(\mu)$ is small being
in contradiction with a calculable evolution. For this reason, if the naive quark
model works at all it is only at very low values of $\mu$ where the evolution is
complete, i.e. where $\alpha_S(\mu)\sim 1$. But then the theoretical accuracy of
the perturbative OPE coefficients is not good because it is governed by the same
$\alpha_S(\mu)$. 

So we employed the following strategy: let us use the  naive quark model for
hadrons  together with a factorization somewhere at small  $\mu$  but let us
allow  adjustment of the OPE coefficients from phenomenological fits. It does not
involve many parameters. Predominantly three coefficients, $c_1$, $c_4$ and
$c_5$, plus a few nonfactorizable matrix elements in hyperon decays determine the
bulk of nonleptonic amplitudes. New relations arising from such a fit  are in
good agreement with experimental data.

\subsection{$\Delta I=3/2$ transitions }
Let us start with the decay $K^+ \to \pi^+ \pi^0$. It is a $\Delta I=3/2$
transition and its amplitude is determined by the matrix element of ${\cal O}_4$,
\begin{equation}
M(K^+ \to \pi^+ \pi^0)=\sqrt{2} \, G_F V_{us}^* V_{ud}\,\langle\pi^+ \pi^0 |c_4{\cal
O}_4|K^+\rangle\ . 
\end{equation}
In the valence quark model the factorization of this matrix element is visible from
the Feynman diagrams presented in Fig.~3. 
\begin{figure}[h]
\epsfxsize=12cm
\centerline{\epsfbox{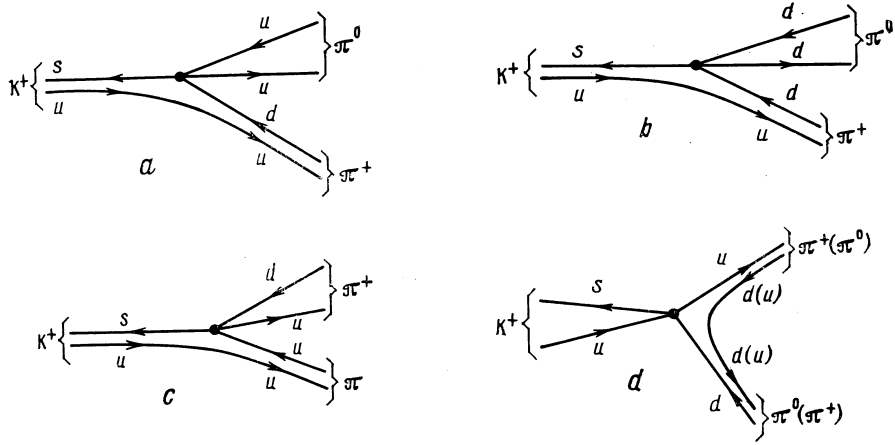}}
\caption{Quark diagrams for $K$ meson decays}
\label{Kdec}
\end{figure}
 
Consider, for instance, the diagram $a$,
\begin{eqnarray}
M_4^a\!\!\!\!\!\!&=&\!\!\!\!\langle \pi^0 |\bar u_L \gamma_\mu u_L|0\rangle
\langle
\pi^+|\bar s_L
\gamma^\mu d_L|K^+\rangle\!\! +\!\!
\langle \pi^0 |(\bar u_L)_i \gamma_\mu (u_L)^j|0\rangle \langle \pi^+|(\bar
s_L)_j
\gamma^\mu (d_L)^i|K^+\rangle\nonumber\\
&=&\!\!\frac{4}{3}\langle \pi^0 |\bar u_L \gamma_\mu
u_L|0\rangle \langle \pi^+|\bar s_L
\gamma^\mu d_L|K^+\rangle\ ,
\end{eqnarray}
where out of three terms entering the definition of ${\cal O}_4$, the first one 
factorizes, the second  factorizes after  Fierz transformation, and the third does not
contribute. The matrix elements in Eq.~(13) are known from semileptonic 
 $\pi\to \mu \nu$ and $K\to \pi e \nu$ transitions,
\begin{eqnarray}
&&\langle \pi^0 |\bar u_L \gamma_\mu u_L|0\rangle
=-\frac{if_\pi}{2\sqrt{2}}q_\mu\,,\qquad f_\pi=0.95\, m_\pi\,,\nonumber\\
&&\langle \pi^+|\bar s_L \gamma_\mu d_L|K^+\rangle=
-\frac 1 2 \left[(p+q_+)_\mu f_+ + (p-q_+)_\mu f_-\right]\,.
\end{eqnarray}
Accounting
for all diagrams in Fig.~3 in a similar way we get
\begin{equation}
M(K^+ \to \pi^+ \pi^0)=ic_4\, G_F V_{us}^* V_{ud}\, m_K^2  f_\pi\,.
\end{equation}
Comparing it with the experimental value
\begin{equation}
\left|M(K^+ \to \pi^+ \pi^0)\right|_{\rm exp}=0.05 \, G_F \, m_K^2 m_\pi
\end{equation}
we find 
\begin{equation}
c_4\approx 0.25\,,
\label{c4}
\end{equation}
what is about 1.6 times less that the theoretical estimate of $c_4$. 

The consistency check comes from the $\Delta I=3/2$ hyperon decays.
In Fig.~4 quark diagrams for the $\Lambda\to p \pi^-$ decay are presented.
\begin{figure}[h]
\epsfxsize=12cm
\centerline{\epsfbox{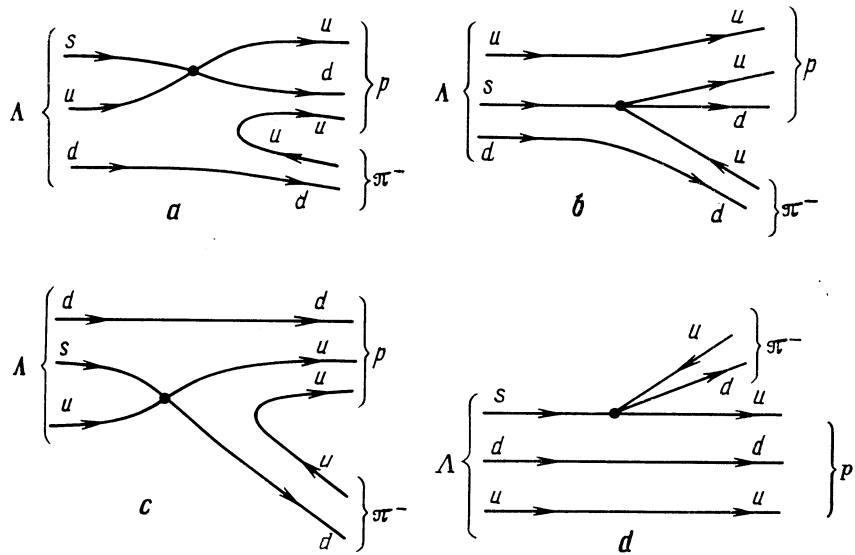}}
\caption{Quark diagrams for hyperon decays}
\label{Ldec}
\end{figure}
\noindent
The symmetry of wave functions and operators under permutations of color indices
(first discussed by Pati and Woo\,\cite{PW})
is important for the analysis, Namely, the operator ${\cal O}_1$ is antisymmetric
under permutation of quark color indices, operators ${\cal O}_{2,3,4}$ are
symmetric, and operators ${\cal O}_{5,6}$ have no specific symmetry, in these
quarks  differ by their helicities. Baryon wave functions are antisymmetric in color,
so  only antisymmetric operators survive in diagrams $a,b,c$. For the 
symmetric $\Delta I=3/2$ operator ${\cal O}_4$ only the diagram 
$c$ remains. For this diagram  factorization takes place,
\begin{eqnarray}
\langle \pi^- p|{\cal O}_4^\dagger|\Lambda\rangle&=&\!\!\frac{1}{\sqrt{2}}\langle
\pi^0 n|{\cal O}_4|\Lambda\rangle=\frac 4 3 \langle \pi^-| \bar d_L \gamma_\mu
u_L|0\rangle \langle p|\bar u_L \gamma^\mu s_L|\Lambda\rangle\nonumber\\
&\approx&\!\! -\frac{i}{\sqrt{6}}f_\pi q_\mu \,\bar u_p \left(\gamma^\mu 
 +\frac 5 9 \,g_A \gamma^\mu\gamma_5\right) u_\Lambda\,.
\end{eqnarray}

Using the value~(\ref{c4}) for the coefficient $c_4$ we get predictions for the
$\Delta I=3/2$  hyperon decay amplitudes. They are collected in the Table 1,
where the $s$ and $p$ wave amplitudes $A$ and $B$ are defined by
\begin{equation}
M=-i G_F m_\pi^2 \,\bar u_f \left(A+B\gamma_5\right) u_i
\end{equation}
\begin{table}[h]
\begin{center}
\begin{tabular}{|c|c|c|c|c|}
\hline
\vspace*{-0.2cm}
~&~&~&~&~\\
{Decay} & {\mbox{$A_{\rm theor}$}} & { 
\mbox{ $A_{\rm exper}$}} &{ \mbox{  $B_{\rm theor}$}} &
{\mbox{  $B_{\rm exper}$}}\\[0.1cm]
\hline
\vspace*{-0.2cm}
~&~&~&~&~\\
{ \mbox{$\Lambda^0_- +\sqrt{2}\Lambda^0_0 $}} &{
  ~0.12} &{ ~0.09 $\pm $ 0.03} &{ $ -0.95$} &{
0.66 $\pm $ 0.81}\\[0.1cm]
{ \mbox{$\sqrt{2} \Sigma^+_0 -\Sigma^+_+ +\Sigma^-_- $}}&{ ~0.14}
&{ ~0.22 $\pm
 $  0.09} &{ ~0.49}
  & { 2.7 $\pm $  1.1}\\[0.1cm]
{ \mbox{$\Xi^-_- +\sqrt{2}\Xi^0_0 $}} &{$-0.14$} &{$-0.15
\pm   0.07$} &{ ~0.23}
 &{ 1.7 $\pm $ 2.2}\\[0.1cm]
\hline
\end{tabular}
\caption{$\Delta I=3/2$ amplitudes in hyperon decays}
\label{32table}
\end{center}
\end{table}
We see that the predictions for $s$ waves are in a reasonable agreement with the
data, for $p$ waves the experimental accuracy is too low for a conclusion.

\subsection{$\Delta I= 1/2$ transitions }
The above analysis of color symmetry and factorization in quark diagrams in
Fig.~4 shows that the operators ${\cal O}_1^\dagger$ (diagrams $a$, $b$, and $c$)
and ${\cal O}_{5}^\dagger$ (the diagram $d$) are dominant in hyperon
decays. Matrix elements of  {${\cal O}_5^\dagger$} are factorizable but that of 
${\cal O}_1^\dagger$ are not. 

There exists a combination of amplitudes,
\begin{equation}
M\left(\Xi^-\to \Sigma^- \pi^0\right)=\sqrt{3} \Lambda^0_- -\Sigma^+_0 
+\sqrt{3}\Xi^-_-\,,
\end{equation}
which does not contain { ${\cal O}_1^\dagger$}. Thus, the ratio of $p$ and $s$ waves
for this combination is predicted,
\begin{equation}
\frac{ B\left(\Xi^-\to \Sigma^- \pi^0\right)} 
{A\left(\Xi^-\to \Sigma^- \pi^0\right)}=g_A\, \frac{m_\Xi
  +m_\Sigma}{m_\Xi -m_\Sigma}\approx 25\,.
\end{equation}
Experimentally this value is $33\pm 10$. The uncertainty can be reduced if the 
Lee-Sugawara relation \mbox{$2\Xi^-_- +\Lambda_-^+ -\sqrt{3}\Sigma^+_0=0$} is
used. Then the value is $27.4 \pm 2.5$.  Thus, we see an experimental confirmation
of our description.

The comparison of $s$ wave $A$
\begin{equation}
A\left(\Xi^-\to \Sigma^- \pi^0\right)= \left(c_5+\frac{3}{16} c_6\right)V_{us}V_{ud}^*\,\frac{2}{9}\, \frac{f_\pi 
  m_\pi^2}{m_u+m_d}\, \frac{m_\Xi -m_\Sigma}{m_s- m_u}
\end{equation}
with the corresponding experimental value $-0.51\pm 0.10 $ fixes
\begin{equation}
c_5+\frac{3}{16} c_6\approx -0.25\,,
\end{equation}
which is about four times larger than the theoretical estimate.

The operators ${\cal O}_{5,6}$ are dominant in $K$ decays. The naive quark model,
together with $\sigma$ meson ($m_\sigma\approx 700~{\rm MeV}$) dominance
in the $s$ wave $\pi\pi$ channel, gives for the matrix element
\begin{eqnarray}
\langle \pi^+\pi^- |{\cal O}_5|K_S\rangle &=& i\, \frac{2\sqrt{2}}{9}\, 
\frac{f_\pi   m_K^2
  m_\pi^2}{(m_u+m_d)\,m_s}\left[\frac{f_K}{f_\pi}\,\frac{1}{m_\sigma^2
    -(q_+ +q_-)^2} -1\right]\nonumber\\[0.2cm]
&\approx &
 i\, \frac{2\sqrt{2}}{9}\, 
\frac{f_\pi   m_K^2
  m_\pi^2}{(m_u+m_d)\,m_s}\left[\frac{f_K}{f_\pi}-1 +
  \frac{f_K}{f_\pi}\,\frac{m_K^2}{m_\sigma^2} \right]\,.
\end{eqnarray}
The result for the amplitude of $K_S\to \pi^+\pi^-$ decay
\begin{equation}
\langle \pi^+\pi^- |H^{\rm eff}|K_S\rangle = (0.85 +0.20)\, i\, G_F \,m_K^2
m_\pi\,,
\end{equation}
where the first number comes from ${\cal O}_{5,6}$ and the second from ${\cal
O}_{1-4}$, matches the experimental value.

\section { Further developments}
\subsection{Limit of large  $N_{\rm color}$, chiral loops}
As we see in the approach presented above the main uncertainty comes from the
range of momenta of the order of hadronic scales. To make a consistent analysis in
this range Bardeen, Buras, and Gerard suggested  to use a description in terms of
chiral meson dynamics\,\cite{BBG}. It means that QCD is used to fix the effective
Hamiltonian  at $\mu$ below charm threshold but large enough to have $\alpha_S
(\mu)\ll 1$, particularly, they choose $\mu\sim 1 {\rm GeV}$. Matrix elements of
this effective Hamiltonian which accounts for momenta  below 1 GeV are calculated
with  mesonic loops instead of quark loops. The use of chiral meson loops can be
parametrically justified in the limit of large number of colors $N_c$. 

Notice that penguin operators appear due to mixing with the left-handed
operators ${\cal O}_{1-4}$, this mixing is suppressed as $1/N_c$. In the dual
mesonic picture it is hadronic vertices which bring in $1/N_c$. The smallness of 
$1/N_c$ emphasizes once more the necessity of a large hadronic scale in the
problem, it was $m_\pi^2/m_q$ in the naive quark model. In chiral dynamics 
$\Delta I=1/2$ amplitudes also come out enhanced what confirm the penguin
explanations of  $\Delta I =1/2$. The chiral loops also improve predictions
for $\Delta I=3/2$ transitions providing an additional suppression. 

\subsection
{ Penguin decays of $B$ mesons}
We mentioned that ``magnetic'' penguins, i.e. the $d=5$ operators of the kind
given  by Eq.~(4) are of particular importance for B decays (see review\,\cite{LSS}).
The Feynman diagram for $b\to s g (\gamma)$ transitions is presented in Fig.~5.
\begin{figure}[h]
\epsfxsize=10cm
\centerline{\epsfbox{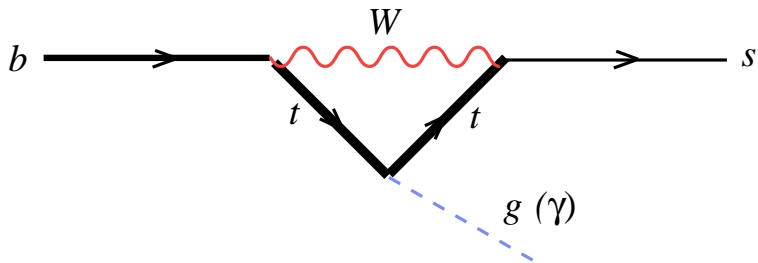}}
\caption{Magnetic penguins for $b\to s g (\gamma)$ transitions }
\label{bsg}
\end{figure}
In difference with strange particle decays there is no suppression due to 
 a small quark mass ($m_b$ enters instead of $m_s$). From the diagram is seen that
``magnetic'' penguins are sensitive to heavy quarks in the loop and, for this
reason, to possible deviations from the Standard Model prediction,
 \begin{equation}
{\rm Br}(b\to s\gamma)= (3.5\pm 0.3)\cdot10^{-4}
\end{equation}
The inclusive rate of $B\to X_s \gamma$ decays was measured by CLEO and ALEPH
collaborations,\\
~\\
\begin{tabular}{rcr}
{ {\rm CLEO:}} &\mbox{$ {\rm Br}(B\to X_s\gamma)=$}&\mbox{$ (2.32\pm 0.57
\pm 0.35)\cdot10^{-4}$}\\[0.2cm]
{ {\rm ALEPH:}}& & \mbox{$(3.11\pm 0.80 \pm
0.72)\cdot10^{-4}$}
\end{tabular}

\vspace{0.4cm}

No deviation was observed. 
This CLEO experiment was recognized at this meeting, Ed Thorndike received the
1999 Panofsky Prize.

\subsection
{ Direct $CP$ violation in $K$ decays }
At this meeting Robert Tschirhart reported a new
measurement of direct $CP$ violation in $K$ decays by KTeV
collaboration\,\cite{ktev}
\begin{equation}
\frac{\epsilon^\prime}{\epsilon}= (2.8\pm 0.41)\cdot 10^{-3}\,,
\end{equation}
which is in agreement with the CERN NA31 experiment but different from the
previous measurement in the Fermilab E731 experiment. 

The direct $CP$ violation is a crucial test of the Standard Model.
Does the result fit Standard Model?  In his presentation Tschirhart defined an
answer to this question as debatable. Indeed, in the recent work\,\cite{BGJLS}
authors claim that the Standard Model leads to the predictions substantially (few
times) lower then the experimental result (27), and only for extreme values of
input parameters the theory can be consistent with the data.

My statement on the issue is that our theory of nonleptonic decays
naturally predicts $\epsilon^\prime/\epsilon$ consistent with Eq.~(27).
Leaving detail for discussion elsewhere\,\cite{VV}, let me
make few remarks.

Actually, relatively large values of $\epsilon^\prime/\epsilon$ within such
approach was obtained long ago by Gilman and Wise\,\cite{GW1}, and by
Voloshin\,\cite{Voloshin}, but then the approach was unfairly abandoned. The
argumentation was, see e.g. Ref.\,\cite{BGJLS}, that the operator ${\cal O}_5$ enters
with the small coefficient at $\mu=m_c$ or $\mu=1\, GeV$, so other operators are
important. This criticism, however,  is not relevant to the dominance of the
operator ${\cal O}_5$  in the {\em low} normalization point where factorization
takes place.

In one loop approximation direct CP violation shows up as an imaginary part of the
coefficient $c_5$
$$
{\rm Im}\, c_5 (m_c)= 
  \frac{{\rm Im}(V_{cs}^*V_{cd})}{V_{us}^*V_{ud}}\,\frac{\alpha_S(m_W)}{12\pi}\ln
    \frac{m_W^2}{m_c^2}\approx 0.12\, {\rm Im}(V_{cs}^*V_{cd})
$$
due to the diagram of Fig.~1 with the $c$ quark in the loop ($t$ quark contribution
is small).  In difference with CP even part, i.e. ${\rm Re}\, c_5$, which comes from
virtual momenta between $\mu$ and $m_c$ (thanks to the GIM cancellation), 
the CP odd part,
${\rm Im}\, c_5$,  comes from a larger range between $m_c$ and $m_W$. The
corrections due to logarithmic evolution   can be simply accounted for, they increase
the value of ${\rm Im}\, c_5(m_c)$. Additionally  ${\rm Im}\, c_5(\mu)$ is
increased by an evolution down to the normalization 
$\mu$ where $\alpha(\mu)\sim 1$ and factorization is applied.

Accounting for the phenomenological value (23) of CP even part we find
$\epsilon^\prime /\epsilon$ in ballpark of the data. Besides confirmation of
the quark mixing nature of CP violation in the Standard Model this serves,
somewhat surprisingly, as one more confirmation of our mechanism for
$\Delta I=1/2$ enhancement. It is nice to get such a surprise on the eve of the
Sakurai Prize.

\section
{Conclusions}
Summarizing the development of twenty four years I cannot, unfortunately, say
that theoretical understanding of $\Delta I=1/2$ is very much advanced.
Progress of last years in technically difficult calculations of higher loops
corrections to OPE coefficients is not crucial when the effect comes  from large
numbers in matrix elements. As we discussed above the enhancement of $\Delta
I=1/2$ reflects the existence of the large momentum scale in light hadrons.
In case of glueballs the scale could be even larger, as it discussed in the next talk 
by Valya Zakharov\,\cite{VZ}.
Moreover, the momentum scale in light hadrons  related with this
enhancement is so large that treatment of the $c$ quark as heavy becomes
questionable.  In this sense the $c$ quark is not heavy enough (let me remind, in
passing, that a low upper limit for its mass was theoretically found\,\cite{KhVG} in
the Standard Model before the experimental discovery of the $c$ quark). 

Thus, we still have some distance to go so I finish by the sentence:

\begin{center}
{\bf Penguins spread out but have not landed yet.}
\end{center}

\section*
{Acknowledgments}
My great thanks to Valya Zakharov and Misha Shifman for a pleasure of our long
term collaboration. I am grateful to the American Physical Society for the honor
to be a recipient of the Sakurai Prize. My appreciation to colleagues, collaborators
and friends in theory groups of Budker Institute and ITEP, especially to B.L.~Ioffe,
I.B.~Khriplovich, I.I~Kogan, V.N.~Novikov,  L.B.~Okun, E.V.~Shuryak, A.V.~Smilga,
V.V.~Sokolov, and M.B.~Voloshin.

This work is supported in part by DOE under the grant number
DE-FG02-94ER40823.


\section*{References}
\begin{thebibliography}{99}
\bibitem{Sch} 
J.~Schwinger, Phys.~Rev.~Lett. {\bf 12}, 630 (1964).
\bibitem{KW}
K.~Wilson, Phys.~Rev. {\bf 179}, 1499 (1969).
\bibitem{PP}
A.Z. Patashinskii and V.L. Pokrovskii, {\it Fluctuation Theory of Phase
Transitions}, Pergamon Press, 1979.
\bibitem{GL}
M.K.~Gaillard and B.~Lee, Phys.~Rev.~Lett. {\bf 33}, 108 (1974).
\bibitem{AM}
G.~Altarelli and L.~Miani, Phys.~Lett. {\bf B52}, 351 (1974).
\bibitem{GIM}
S.L.~Glashow, J.~Illiopoulos, and L.~Miani, Phys.~Rev. {\bf D2}, 1285 (1970).
\bibitem{Leut}
H.~Leutwyller, Phys.~Lett. {\bf B48}, 45 (1974); Nucl.~Phys. {\bf B76}, 413 (1974).
\bibitem{MGM}
M.~Gell-Mann, Elementary Particles, Oppenheimer Lectures, Institute for Advanced
Study  preprints 75-0978, 75-0979.
\bibitem{VZS}
A.I.~Vainshtein, V.I.~Zakharov, and M.A.~Shifman, \\
Pisma Zh.~Eksp.~Teor.~Fiz. {\bf 22}, 123 (1975) [JETP Lett. {\bf 22} 55 (1975)]; \\
Nucl.~Phys. {\bf B120}, 316 (1977);
\\Zh. Eksp. Teor. Fiz. {\bf 72}, 1275 (1977) [Sov. Phys. JETP {\bf 45} 670, (1977)].
\bibitem{EGNR}
J.~Ellis, M.K.~Gaillard, D.V.~Nanopoulos, and S.~Rudaz,\\
Nucl.~Phys. {\bf B131}, 285 (1977), (E) {\bf B132} 541 (1978) 
\bibitem{MS}
M.~Shifman, {\it ITEP Lectures in Particle Physics}, hep-ph/9510397.
\bibitem{GW}
F.J.~Gilman and M.B.~Wise, Phys.~Rev. {\bf D20}, 2392 (1979).
\bibitem{BBL}
G.~Buchalla, A.J.~Buras, and M.E.~Lautenbacher,
Rev.~Mod.~Phys. {\bf 68}, 1125 (1996).
\bibitem{PW}
J.C.~Pati and C.H.~Woo,  Phys.~Rev. {\bf D3}, 2920 (1972).
\bibitem{BBG}
W.A.~Bardeen, A.J.~Buras, and J.-M.Gerard,
 Phys.~Lett. {\bf B192}, 138 (1987); Nucl.~Phys. {\bf B293}, 787 (1987).
\bibitem{LSS}
K.~Lingel,  T.~Skwarnicki, and J.G.~Smith, {\em Penguin Decays of B Mesons}.
hep-ex/9804015.
\bibitem{ktev}
KTeV collaboration, {\em
Observation of Direct CP Violation in $K_{S,L}$ to $\pi \pi$ Decays},
hep-ex/9905060.
\bibitem{BGJLS}
S.~Bosch {\em et al}, {\em Standard Model Confronting New Results for
epsilon'/epsilon}, hep-ph/9904408.
\bibitem{VV}
A.~Vainshtein and M.~Voloshin, in preparation.
\bibitem{GW1}
F.J.~Gilman and M.B.~Wise,
Phys. Lett. {\bf 83B}, 83 (1979).
\bibitem{Voloshin}
M.B.~Voloshin, {\em
Introduction To Quantum Chromodynamics. Effects Of Strong Interactions In
Nonleptonic Weak Decays}, Preprint ITEP-22-1981.
\bibitem{KhVG}
A.I.~Vainshtein and I.B.~Khriplovich,
Pisma Zh.~Eksp.~Teor.~Fiz. {\bf 18}, 141 (1973);\\
M.K.~Gaillard and B.W.~Lee, Phys. Rev. {\bf D10}, 897 (1974)
\bibitem{VZ}
V.I.~Zakharov, {\em Gluon Condensate and Beyond}, The 1999 Sakurai Prize Lecture
at the Centennial Meeting of the APS, hep-ph/9906xxx .


\end{thebibliography}
\end{document}